\documentclass[preprint2]{aastex}
\usepackage[latin1]{inputenc}
\usepackage{graphicx}
\graphicspath{ {./} }
\usepackage{amsmath}

\DeclareMathOperator*{\argmin}{arg\,min}

\slugcomment{}

\shorttitle{}
\shortauthors{Guennou et al.}

\begin{document}


\title{On the Accuracy of the Differential Emission Measure Diagnostics of Solar Plasmas. Application to AIA / {\it \textbf{SDO}}. Part II: Multithermal plasmas.\\}


\author{C. Guennou\altaffilmark{1}, F. Auch\`ere\altaffilmark{1}, E. Soubri\'e\altaffilmark{1} and K. Bocchialini\altaffilmark{1}}
\affil{Institut d'Astrophysique Spatiale, B\^atiment 121, CNRS/Universit\'e Paris-Sud, UMR 8617, 91405 Orsay, France}
\email{chloe.guennou@ias.u-psud.fr}

\author{S. Parenti\altaffilmark{2}}
\affil{Royal Observatory of Belgium, 3 Avenue Circulaire, B-1180 Bruxelles, Belgium}

\and

\author{N. Barbey\altaffilmark{3}}
\affil{SAp/Irfu/DSM/CEA, Centre d'\'etudes de Saclay, Orme des Merisiers, B\^atiment 709, 91191 Gif sur Yvette, France}

\affil{Accepted for publication in The Astrophysical Journal Supplements 2012 September 5.}



\begin{abstract}

The Differential Emission Measure (DEM) analysis is one of the most used diagnostic tools for solar and stellar coronae. Being an inverse problem, it has limitations due to the presence of random and systematic errors. We present in theses series of papers an analysis of the robustness of the inversion in the case of AIA/SDO observations. We completely characterize the DEM inversion and its statistical properties, providing all the solutions consistent with the data along with their associated probabilities, and a test of the suitability of the assumed DEM model.

While Paper I focused on isothermal conditions, we now consider multi-thermal plasmas and investigate both isothermal and multithermal solutions. We demonstrate how the ambiguity between noises and multi-thermality fundamentally limits the temperature resolution of the inversion. We show that if the observed plasma is multi-thermal, isothermal solutions tend to cluster on a constant temperature whatever the number of passbands or spectral lines. The multi-thermal solutions are also found to be biased toward near isothermal solutions around 1 MK. This is true even if the residuals support the chosen DEM model, possibly leading to erroneous conclusions on the observed plasma. We propose tools to identify and quantify the possible degeneracy of solutions, thus helping the interpretation of DEM inversion.  
\end{abstract}



\section{Introduction}
\label{intro}

A convenient approach to study the thermal structure of the solar and stellar outer atmospheres is the Differential Emission Measure (DEM) formalism. The DEM $\xi(T_e)$ is a measure of the amount of emitting plasma along the Line Of Sight (LOS) as a function of the electron temperature $T_e$. However, the intrinsic difficulties involved in this inverse problem lead to many complications in its inference, making its interpretation ambiguous. The central point of these series of papers is to provide new tools to systematically and completely characterize the DEM inversions and assist the DEM interpretation. Using the technique developed for this purpose, exhaustively described in ~\citep[][hereafter Paper I]{guennou2012}, it is possible to determine and to compare the DEM diagnostic capabilities of given instruments. With only three bands, the previous generation of imaging telescopes were shown not to be well-suited to DEM analysis~\citep[e.g.][]{schmelz2009}. The situation has changed with the availability of new multi-band instruments such as the Atmospheric Imaging Assembly (AIA) telescope \citep{lemen2012} on board the \textit{Solar Dynamics Observatory} (SDO). Applying our technique to AIA, we show the increased robustness of the inversion for isothermal plasmas, but we also found intrinsic biases if the observed plasma is multithermal.      

Using the notations of Paper I, the DEM is defined as
\begin{equation}
\xi(T_e)=\overline{n_e^2}(T_e)\,dp/d(\log T_e),
\label{eq:dem}
\end{equation}
where $\overline{n_{e}^2}$ is the square electron density averaged over the portions $dp$ of the LOS at temperature $T_{e}$ \citep{craig1976}. The observed intensity in the spectral band $b$ of an instrument can be expressed as a function of the DEM $\xi$ as follows 
\begin{equation}
I_b=\frac{1}{4\pi}\int_{0}^{+\infty} R_b(n_e, T_e)\, \xi(T_{e})\, d\log T_{e}.
\label{eq:intensity_dem}
\end{equation}
Details and references about the DEM formalism can be found in Paper I.
Given a set of observations in $N$ different bands, the DEM can be in principle inferred by reversing the image acquisition process described by Equation~\ref{eq:intensity_dem}. However, solving for the DEM proved to be a considerable challenge. The complications involved in its derivation are one of the reasons for controversial results regarding the thermal structure of the corona. For example, the DEM is one of the method used to derive the still debated physical properties of plumes~\citep{wilhelm2011}. Also, while heating models predict different thermal structures for coronal loops depending on the processes involved \citep[e.g.][]{klimchuk2006, reale2010}, DEM analyses provided ambiguous answers~\citep[][]{schmelz2009}. Several studies \citep{feldman1998, warren1999, landi2008} suggest the ubiquitous presence of isothermal plasma in the quiet corona. If confirmed, these results would challenge many theoretical models, but the reliability of the temperature diagnostics used has been questioned~\citep[e.g.][]{landi2012}. Part of this state of facts is due to technical issues, such as the difficulty to subtract the background emission~\citep{terzo2010, aschwanden2008} or the possible spatial and temporal mismatch of the structures observed in different bands and/or different instruments. But the fundamental limitations of the DEM inversion clearly play a role.

Despite the many proposed inversion schemes, rigorously estimating levels of confidence in the various possible solutions given the uncertainties remains a major difficulty. In this work we propose a strategy to explore the whole parameter space, to detect the presence of secondary solutions, and to compute their respective probabilities. We illustrate the method by characterizing the inversion of simulated observations of the AIA telescope. This approach allows us to quantify the robustness of the inversion by comparing the inverted DEM to the input of the AIA simulations. The main drawback is the limitation to simple DEM forms (i.e. described by a small number of parameters), but the results provide important insights into the properties of more generic DEM inversions. The method can also easily be applied to any instrument, broad band imagers and spectrometers alike.

Paper I was dedicated to the analysis of isothermal plasmas, and showed the existence of multiple solutions in the case of a limited number of bands. Notwithstanding, the statistical method developed in this work (which provides the respective probabilities of each secondary solution), enables to properly interpret the solutions as consistent with several plasma temperatures. The use of the six AIA coronal bands was shown to increase the thermal diagnostic capabilities, providing a robust reconstructions of isothermal plasmas. The detailed analysis of the squared residuals showed that the DEM complexity is limited by the redundancy of information between the bands. 

\begin{figure*}
\epsscale{1.7}
\begin{center}
\plotone{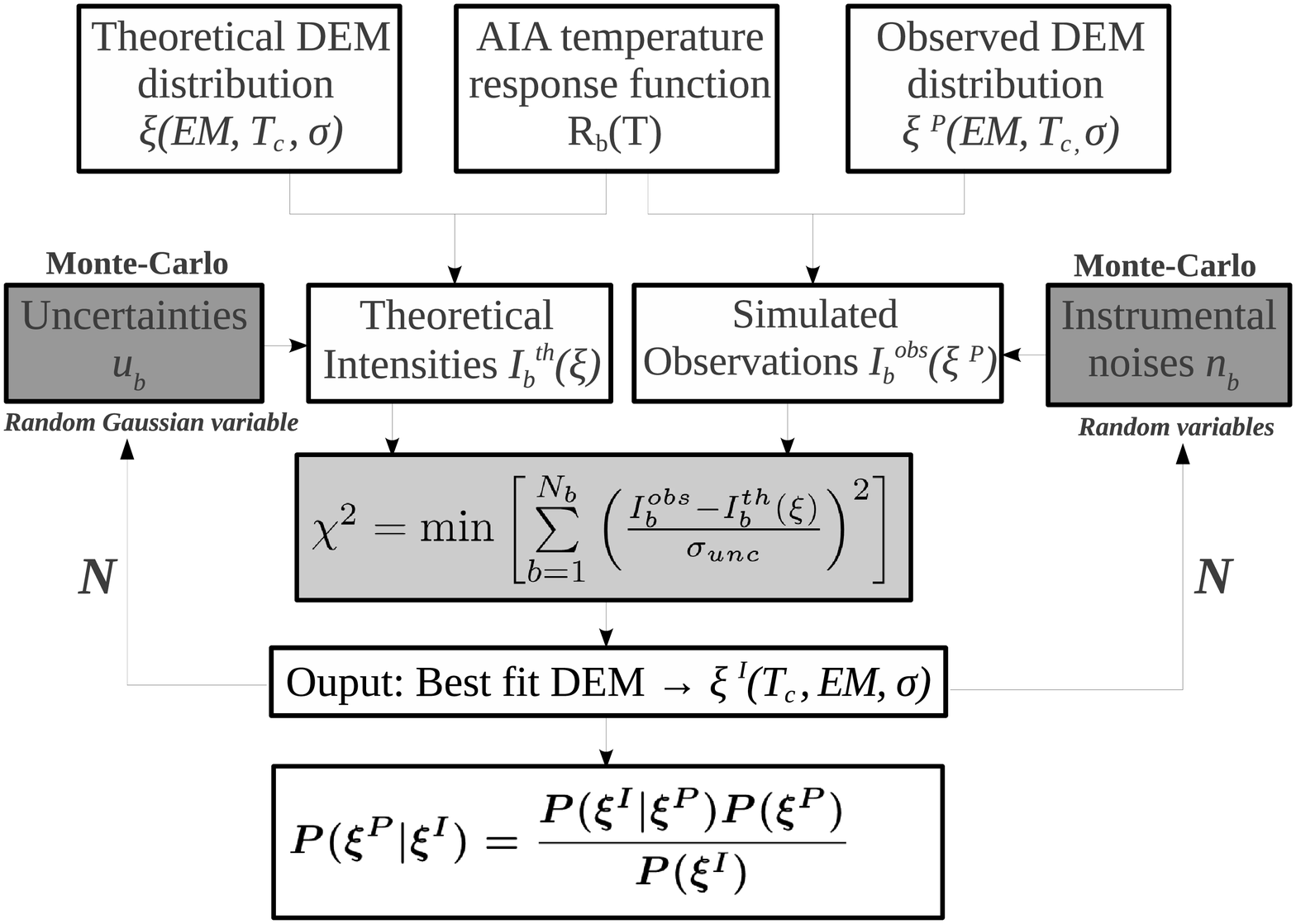}
\end{center}
\caption{Summary of the technique used in this work. The AIA temperature response functions are computed using CHIANTI 7.0. The simulated observations $I_b^{obs}$ are calculated assuming a simple DEM model $\xi^P$ (isothermal, Gaussian or top hat), and adding the instrumental random perturbations and systematic errors. The theoretical intensities  $I_b^{th}$ are computed assuming a DEM model $\xi$, for a large range of parameters. The inverted DEM $\xi^I$ is evaluated by least-square minimization of the distance between the theoretical intensities and the simulated observations. Using a Monte-Carlo scheme, $N$ realizations of the random and systematic errors $n_b$ and $s_b$ are computed. For a given set of parameters $(EM^P, T_c^P, \sigma^P)$ of the plasma DEM $\xi^P$, the quantity $P(\xi^I|\xi^P)$ is then evaluated. The probability $P(\xi^P|\xi^I)$ is then derived using Bayes' theorem.\label{overview}}
\end{figure*}

We now generalize the method to DEM models able to describe a great variety of plasma conditions, from isothermal to broadly multithermal. A summary of the technique is given Figure \ref{overview} and its implementation is exhaustively described in Paper I. The core of our method resides in the probabilistic interpretation of the DEM solutions. The DEM $\xi^I$ solution of the inversion is the one that minimizes a criterion $C(\xi)$ defined as the distance between the intensities $I_b^{obs}$ observed in $N_b$ bands and the theoretical ones $I_b^{th}$. We write
\begin{equation}
\begin{split}
C(\xi) = & \sum_{b=1}^{N_b} \left( \frac{I_b^{obs}(\xi^P) - I_b^{th}(\xi)}{\sigma_b^u}\right)^{\rm 2}\\
\xi^I = & \argmin_{\xi} C(\xi)\\
\chi^2 =  &\min C(\xi)
\end{split}
\label{eq:criterion}
\end{equation}
where $\sigma_b^u$ is the standard deviation of the uncertainties and $\chi^2$ represents the residuals. Using Monte-Carlo simulations of the systematic and random errors, we compute $P(\xi^I|\xi^P)$ the conditional probability to obtain a DEM $\xi^I$ knowing the plasma DEM $\xi^P$. Both types of errors are modelled as Gaussian random variables with a 25\% standard deviation (see Section 2.3.3 of Paper I). Then, using Bayes' theorem we obtain $P(\xi^P|\xi^I)$, the probability that the plasma has a DEM $\xi^P$ given the inverted DEM $\xi^I$. Using this latter quantity it is possible to identify the range or multiple ranges of solutions consistent with a set observations. Therefore, even if it is not possible to alleviate the degeneracy of the solutions, it is at least possible to notice and to quantify them. But the computation of the probability $P(\xi^P|\xi^I)$ is practical only if the space of solutions is limited. We therefore restrict the possible solutions to simple forms described by a small number a parameters: Dirac delta function, top hat, and Gaussian. We thus do not to propose a generic inversion algorithm, but using this approach, we were able to completely characterize the behavior of the inversion in well controlled experiments.

All the results presented in the present paper concern multithermal plasmas. In Section~\ref{iso_response}, we analyse the properties of the isothermal solutions in order to determine to what extent it is possible to discriminate between isothermal and multithermal plasmas. Section \ref{multi_response} is then dedicated to the general case of multithermal inversions. Results are summarized and discussed in Section~\ref{discussion}.   

\section{Isothermal solutions}
\label{iso_response}

As already mentioned, a recurring question is that of the isothermality of solar plasmas. The EM loci technique~\citep[e.g.][and references therein]{delzanna2002, delzanna2003} was originally proposed as an isothermality test. The EM loci curves are formed by the set of $(EM, T_e)$ pairs for which the isothermal theoretical intensities exactly match the observations in a given band or spectral line. The isothermal hypothesis can then in principle be ruled out if there is no single crossing point of the loci curves. But a fundamental ambiguity arises from the inevitable presence of measurement errors.

Under the hypothesis of ideal measurements, the observations and the theoretical intensities are both reduced to $I_b^0$. Hence, the isothermal solution for a perfectly isothermal plasma leads to a residual $\chi^2$ exactly equal to 0, and the EM loci curves all intersect at a common point. But, in reality, the presence of errors leads to over or under-estimations of the theoretical and observed intensities, which conducts to a non-zero $\chi^2$. In this case, even though the plasma is isothermal, the EM loci curves do not intersect at a single point, each one being randomly shifted from its original position. But even with perfect measurements, the isothermal hypothesis would also yield a residual greater than zero if the observed plasma is multithermal, because in this case the $I_b^0$ and $I_b^{th}$ are intrinsically different. Therefore, measurement errors can be incorrectly interpreted as deviations from isothermality and {\it vice-versa}, a variety of multithermal plasmas can be statistically consistent with the isothermal hypothesis. The question is then whether or not a perfectly isothermal plasma can be distinguished from a multithermal one and if so, under what conditions.

We thus investigate in this section the isothermal solutions $\xi^I_{iso}$ obtained from simulated observations $I_b^{obs}$ of plasmas of different degrees of multithermality. The adopted statistical approach allows us to quantify to what extent the robustness of the inversion is affected as a function of the degree of multithermality of the plasma. The plasma is assumed to have a Gaussian DEM defined in $\log T_e$ as
\begin{equation}
\begin{split}
\xi_{gau}^P(EM, T_c, \sigma) = & EM\mathcal{N}(\log T_e - \log T_c),\\
\mbox{with } \mathcal{N}(x) = & \frac{1}{\sigma\sqrt{2\pi}} \exp\left(-\frac{x^2}{2\sigma^2}\right)
\end{split}
\label{eq:dem_gauss}
\end{equation}
where $EM$ is the total emission measure, $T_c$ is the central temperature and $\sigma$ is the width of the DEM. It represents a plasma predominantly distributed around a central temperature $T_c$ and can represent both isothermal and very multithermal plasmas.

The theoretical intensities $I_b^{th}$ are limited to the case of isothermal solutions, corresponding to a DEM 
\begin{equation}
\xi_{iso}(EM, T_{c}) = EM\, \delta(T_e - T_c),
\label{eq:dem_iso}
\end{equation}
where $\delta$ is Dirac's delta function and EM is the total emission measure. The solution $\xi^I$ of the inversion process then gives the pair ($T_c^I, EM^I$) that best explains the observations. The reference intensities $I_b^0$ (see the previous Section), used to compute both $I_b^{obs}$ and $I_b^{th}$ have been tabulated once and for all for 500 central temperatures from $\log T_c(\mathrm{K})=5$ to 7.5, 200 emission measures from $\log EM(\mathrm{cm}^{-5}) = 25$ to 33, and for the Gaussian model 80 DEM widths from 0 to 0.8, expressed in unit of $\log T_e$ (see Section 2.3.2 of Paper I).

\subsection{Ambiguity between uncertainties and multithermality}
\label{sec:ambiguity}

\begin{figure*}
\begin{center}
\includegraphics[scale=1]{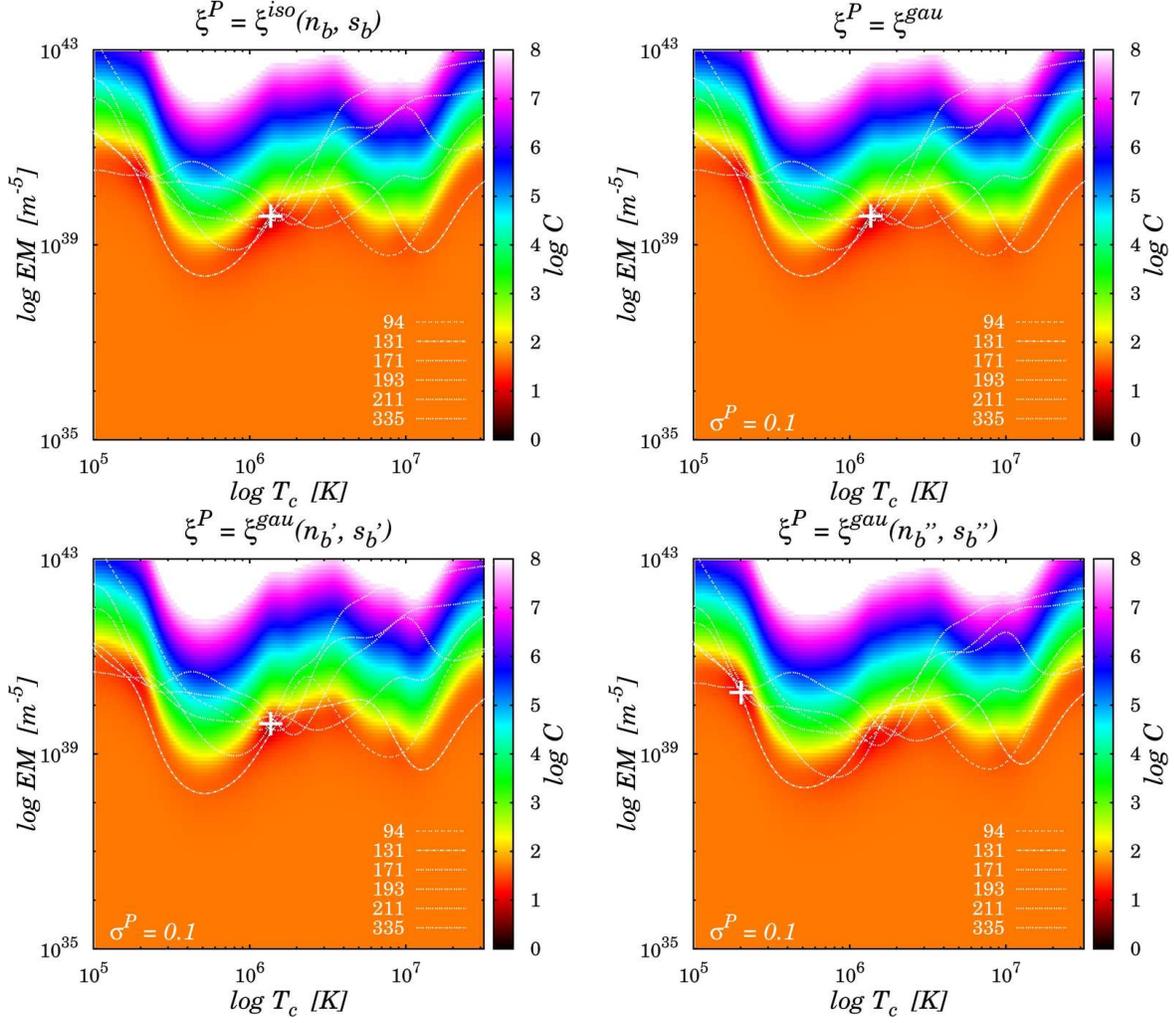}
\end{center}
\caption{Equivalence between uncertainties and multi-thermality. The criterion $C(EM, T_c)$ given by Equation~\ref{eq:criterion}, represented in gray scale and its absolute minimum, marked by a white plus sign. The EM loci curves for each of the six AIA coronal bands are also represented (white lines), superimposed to the criterion. The location of the absolute minimum provides the isothermal solution $\xi^I$ for a plasma having DEMs $\xi^P$ centered on $T_c^P=1.5$~MK, and a total emission measure of $2\times 10^{29}\ \mathrm{cm}^{-5}$. \textit{Top left}: isothermal plasma having a DEM $\xi^P = \xi^{iso}$ in presence of random and systematic errors. \textit{Top right}: Multithermal plasma having a Gaussian DEM $\xi^P = \xi^{gauss}$ with a width $\sigma^P = 0.1\ \log T_e$, without uncertainties. Uncertainties and multithermality both produce comparable deviations of the EM loci curves. \textit{Bottom}: Same as top right, but for two different realizations of the random and systematic errors $n_b$ and $s_b$. } \label{criterion}
\end{figure*}

In Figure~\ref{criterion}, the criterion $C(EM, T_c)$ (Equation \ref{eq:criterion}) is plotted for different plasma configurations. It is represented in gray scale\footnote{All the criteria presented in this paper are available in color on line at ftp.ias.u-psud.fr/cguennou/DEM\_AIA\_inversion/.} and its absolute minimum, corresponding to the solution $\xi^I$, is marked by a white plus sign. The criterion is a function of the two parameters $EM$ and $T_c$ defining the isothermal solutions (Equation (\ref{eq:dem_iso})). The white curves are the EM loci curves, corresponding to the minimum of the contribution of each band to the total criterion. In all panels, the simulated plasmas have DEMs centered on the typical coronal temperature $T_c^P=1.5 \times 10^6$~K, and a total emission measure $EM^P=2\times 10^{29}\ \mathrm{cm}^{-5}$, an active region value guaranteeing signal in all six bands (see Section 3 and Figure 3 in Paper I). The top left panel corresponds to an isothermal plasma ($\sigma=0$ which corresponds to the case presented in Section 3.2 of Paper I). Because of systematics, $s_b$, and instrumental noises, $n_b$, and thus the over or under-estimation of the observed and theoretical intensities $I_b^{obs}$ and $I_b^{th}$, the EM loci curves are shifted up and down along the emission measure axis, with respect to the zero-uncertainties case (i.e. $n_b$ and $s_b = 0$). Therefore, the curves never intersect all six at a single position, thus giving $\chi^2 > 0$. But multithermality has the same effect: the criterion presented on the top right corresponds to a slightly multithermal plasma having a Gaussian DEM width of $\sigma^P$ = 0.1 $\log T_e$. Even though in this case uncertainties have not been added, the loci curves and the absolute minimum, corresponding to the location where the curves are the closest together, are also shifted relative to each other along the emission measure axis and there is no single crossing point. Thus, errors and multithermality can both produce comparable deviations of the observations from the ideal isothermal case, hence the fundamental ambiguity between the two. The two criteria of the bottom panels have been obtained with the same plasma Gaussian DEM distribution, but now in presence of random perturbations. These two independent realizations of the uncertainties illustrate an example of the resulting dispersion of the solutions.

\begin{figure*}
\begin{center}
\includegraphics[scale=1]{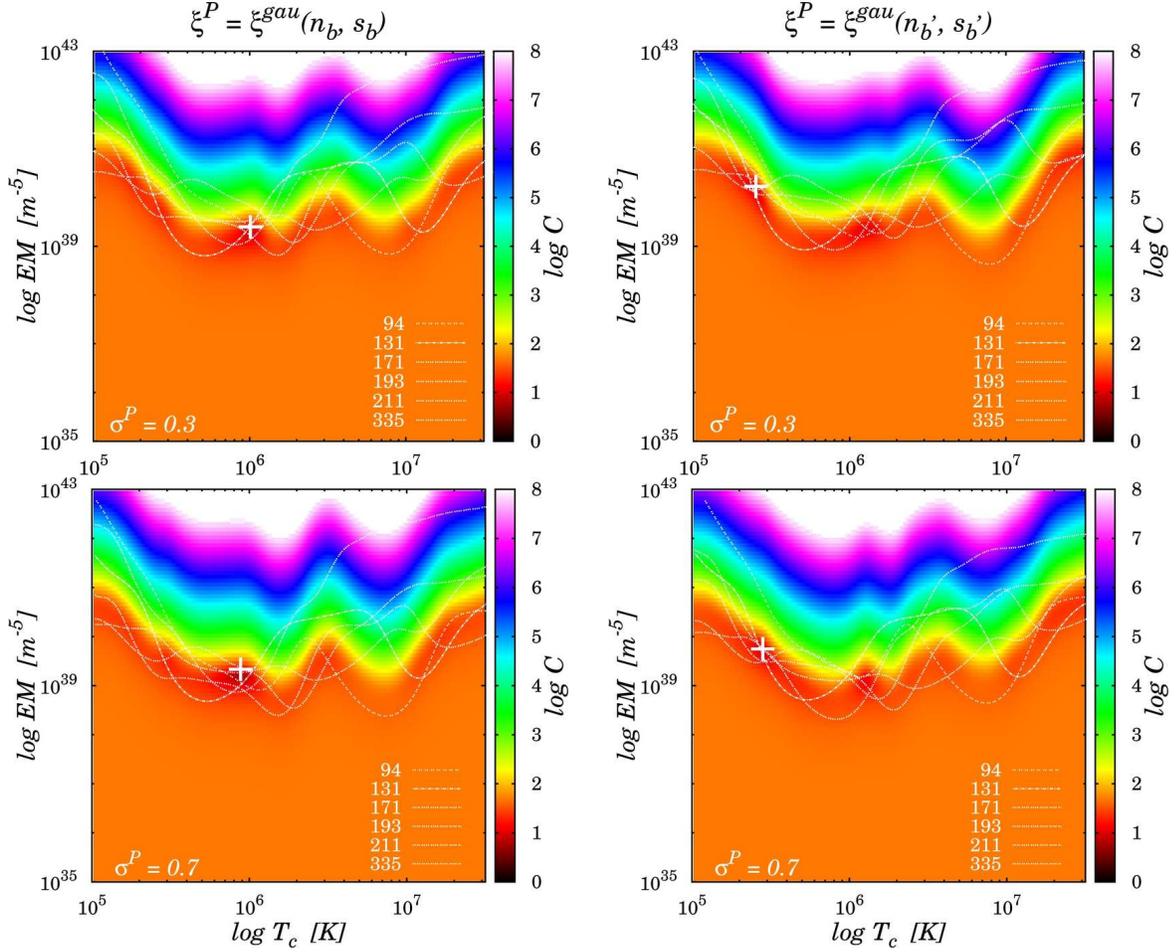}
\end{center}
\caption{Isothermal solutions for different thermal widths. Same as Figure~\ref{criterion} for larger DEMs plasma widths. In the top row the plasma has a Gaussian DEM $\sigma^P=0.3\ \log T_e$, and $\sigma^P=0.7\ \log T_e$  in the bottom row. For each thermal width, two different realizations of the random and systematic errors are presented, showing that the location of the absolute minimum can greatly vary.} \label{fig:criterion_gi}
\end{figure*}

In Figure \ref{fig:criterion_gi}, we increased the DEM width to $\sigma^P = 0.3\ \log T_e$ (top row) and $\sigma^P = 0.7\ \log T_e$ (bottom row) while keeping the same emission measure and central temperature. For each width, the left and right panels show two realizations of the uncertainties. The corresponding configuration of the loci curves and thus the value and position of the absolute minimum can greatly vary, even though the plasma is identical in both cases. It also appears that it exists privileged temperature intervals where the solutions tend to concentrate. This phenomenon is intrinsically due to the shape of the EM loci curves. In the next subsection, we characterize the respective probabilities of occurrence of the various solutions for a wide range of plasma conditions. From the examples presented in Figure \ref{fig:criterion_gi}, we also note that the vertical spread of the loci curves tends to be larger for wider DEMs, which corresponds to larger $\chi^2$ residuals. In section~\ref{sec:interpretation_21}, we will determine up to what DEM width the observations can be considered to be consistent with the isothermal hypothesis.

\subsection{Probability maps}
\label{sec:gauss-iso_maps}

We consider Gaussian DEM plasmas and scan all possible combinations of central temperatures and widths used to pre-compute the reference theoretical intensities. The total emission measure is kept constant at $EM^P = 2\times 10^{29}\ \mathrm{cm}^{-5}$. For each combination of plasma parameters, 5000 Monte-Carlo realizations of the random perturbations $n_b$ and systematics $s_b$ are obtained, leading to 5000 estimates of the observed intensities $I_b^{obs}(\xi^P)$ in each band $b$. The isothermal least-square inversion of these 5000 sets of AIA simulated observations provides as many solutions $\xi^I$, allowing an estimation of $P(\xi^I|\xi^P)$.

\begin{figure*}
\begin{center}
\includegraphics[scale=1]{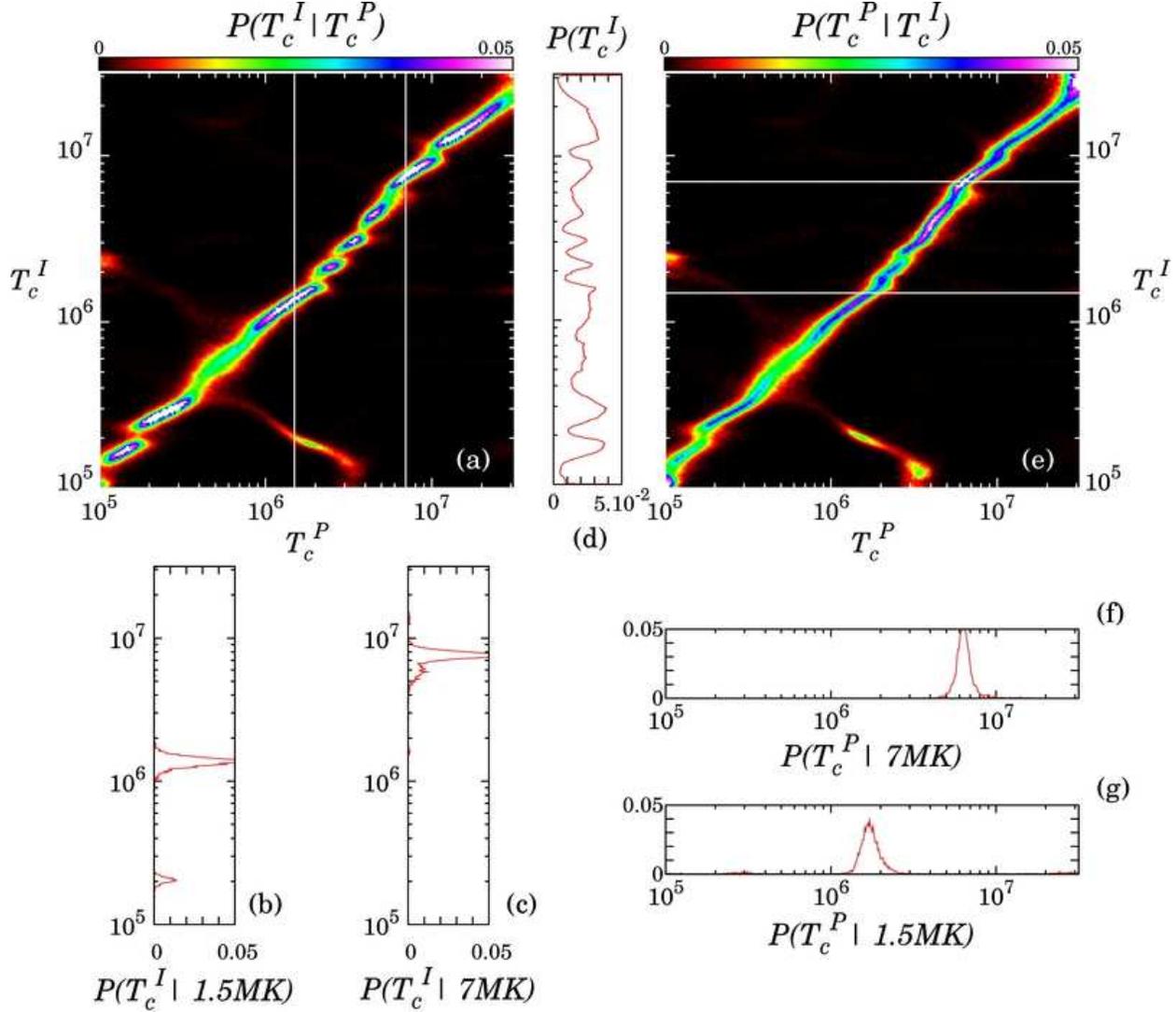}
\end{center}
\caption{Maps of probabilities considering a Gaussian DEM plasma $\xi^P = \xi^{gau}$ having a narrow thermal distribution of $\sigma^P = 0.1 \log T_e$, obtained by 5000 Monte-Carlo realizations of the random and systematics errors $n_b$ and $s_b$, and investigating the isothermal solutions. \textit{(a)}: Probability map $P(T_c^I|T_c^P)$, vertically reading. The central temperature $T_c^I$ resulting from the inversion is presented whatever the total emission measure $EM^I$. \textit{(b)} and \textit{(c)}: Probability profiles of $T_c^I$ for plasma central temperatures $T_c^P = 1.5\times 10^6$ and $7\times 10^6$~K (vertical lines in  panel(a)). \textit{(d)}: Total probability to obtain $T_c^I$ whatever $T_c^P$ (see Section \ref{intro} and Section 2.1 of paper I). \textit{(e)}: \textit{Vice-versa}, probability map $P(T_c^P|T_c^I)$, horizontally reading, inferred by means of Bayes' theorem using $P(T_c^I| T_c^P)$ and $P(T_c^I)$. \textit{(f)} and \textit{(g)}: Probability profiles of $T_c^P$ knowing that the inversion result is, from top to bottom, $7 \times 10^6$ and $1.5\times10^6$~K.}\label{fig:multi_slight_vs_iso}
\end{figure*}

\begin{figure*}
\begin{center}
\includegraphics[scale=1]{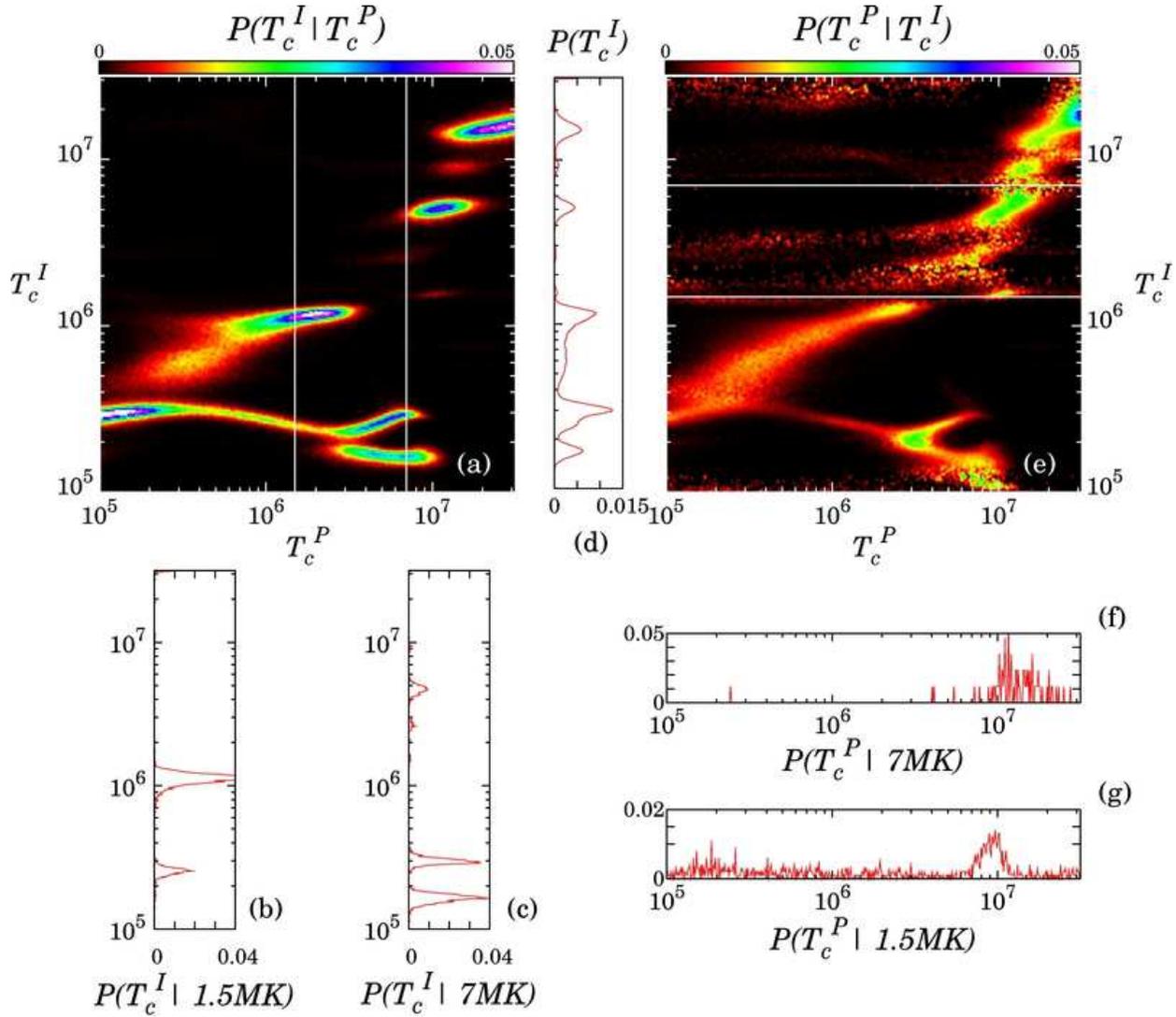}
\end{center}
\caption{Same as Figure \ref{fig:multi_slight_vs_iso}, but with a plasma DEM width is increased to $\sigma^P = 0.3\ \log T_e$. Many perturbations, already visible in the Figure~\ref{fig:multi_slight_vs_iso} are amplified: the distribution is wider, irregular and the diagonal structure disappeared (see panel (a)). The presence of multiples solutions of comparable probabilities is increased for a large range of plasma temperatures $T_c^P$, leading to very different estimated $T_c^I$ from the input $T_c^P$. The probability map $P(T_c^I| T_c^P)$ can help us to properly interpret the inversion result, taking into account the secondary solutions and providing their respective probability.}\label{fig:multi_0.3_vs_iso}
\end{figure*}

\begin{figure*}
\begin{center}
\includegraphics[scale=1]{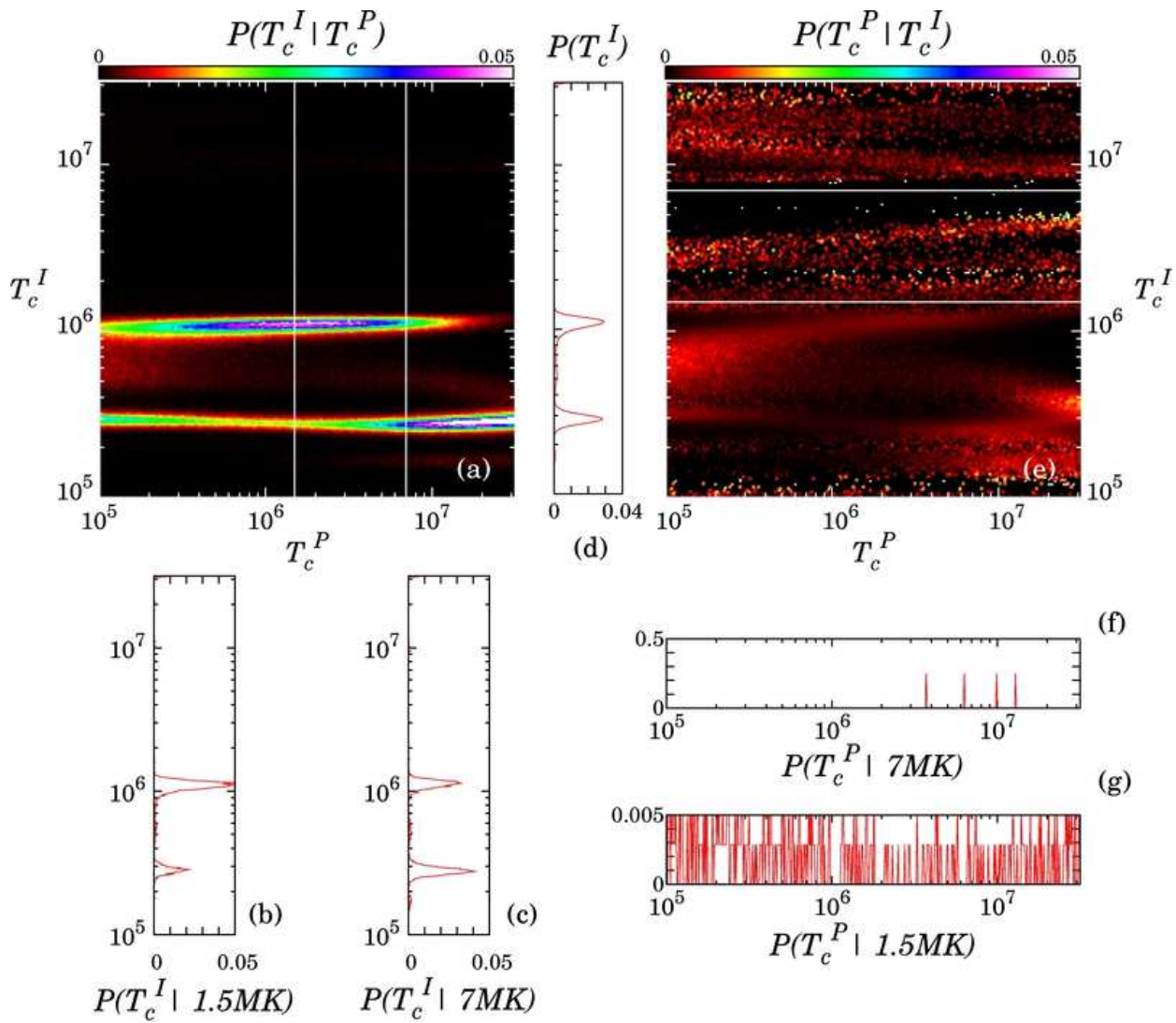}
\end{center}
\caption{Same as Figure \ref{fig:multi_slight_vs_iso}, but in case of the plasma DEM width is increased to $\sigma^P = 0.7\ \log T_e$. The impact on the robustness of the inversion is clearly increased in this case, showing two privileged isothermal solutions $T_c^I$, totally decorrelated from the input $T_c^P$. As a result, no information regarding the central temperature of the DEM can be extracted from the inversion.}\label{fig:multi_0.7_vs_iso}
\end{figure*}

Figures~\ref{fig:multi_slight_vs_iso},~\ref{fig:multi_0.3_vs_iso} and~\ref{fig:multi_0.7_vs_iso} show the resulting temperature probability maps for $\sigma^P=0.1$, 0.3 and 0.7~$\log T_e$ respectively\footnote{The probability maps for 80 widths from $0$ to $0.8\ \log T_e$ are available in color on line at ftp.ias.u-psud.fr/cguennou/DEM\_AIA\_inversion/.}. The isothermal plasma case ($\sigma^P=0$) is shown in Figure 6 of Paper I. In all figures, the probability $P(T_c^I|T_c^P)$ is obtained by reading vertically panels (a), given the probability to find a solution $T_c^I$ knowing the plasma central temperature $T_c^P$. The probability profiles for two specific plasma temperatures, $1.5 \times 10^6$ and $7\times 10^6$~K, are shown in panels (b) and (c). Using Baye's theorem, the probability maps $P(T_c^P|T_c^I)$ of panels (e) are obtained by normalizing $P(T_c^I|T_c^P)$ to $P(T_c^I)$, the probability of obtaining $T_c^I$ whatever $T_c^P$ (panels (d)). In the bottom right panels (f) and (g), we show two example horizontal profiles giving the probability that the plasma has a central temperature $T_c^P$ knowing the inverted temperature $T_c^I$.

In panels (a) and (e) of Figure~\ref{fig:multi_slight_vs_iso}, the most probable solutions are located around the diagonal. However, compared to the case of an isothermal plasma (see Figure 6 of Paper I), the distribution is wider, irregular and deviations from the diagonal greater than its local width are present. As shown by panel (d) and the nodosities in the map (a), the unconditional probability to obtain a result $T_c^I$ is non uniform, meaning that some inverted temperatures are privileged whereas others are unlikely. Comparing with Figure 6 of Paper I, profile (b) shows that the probability of secondary solutions at $T_c^P=1.5\times 10^6$~K is increased with respect to the isothermal case. The apparition of these two solutions are illustrated in the bottom row of Figure~\ref{criterion}. The bottom right panel corresponds to a realization of the errors yielding a solution close to the diagonal, while the bottom left panel of the same Figure illustrates a case where the absolute minimum of the criterion is located at low temperature. Using the map of $P(T_c^P|T_c^I)$ it is however possible to correctly interpret the low temperature solutions as also compatible with $1.5\times 10^6$~K plasma (profile (g)). 

In Figure~\ref{fig:multi_0.3_vs_iso}, the plasma DEM width is increased to $\sigma^P=0.3\ \log T_e$. As a result, the above-described perturbations with respect to the isothermal plasma case are amplified. The diagonal structure has almost disappeared, with discontinuities and reinforced and more diffuse nodosities. Multiple solutions of comparable probabilities are present over large ranges of plasma temperatures and consequently, the estimated $T_c^I$ can be very different from the input $T_c^P$. For example, panel (c) shows that for a $7\times 10^6$~K plasma, the most probable $T_c^I$ is either $1.6\times 10^5$ or $3\times 10^5$~K. The unconditional probability $P(T_c^I)$ of panel (d) is very non uniform, some ranges of estimated temperatures being totally unlikely (e.g. $T_c^I=[1.5\times 10^6, 4\times 10^6]$~K) while others are probable for large intervals of $T_c^P$ (e.g. $T_c^I=3\times 10^5$~K or $10^6$~K). However, despite the jaggedness of $P(T_c^I|T_c^P)$, the map of $P(T_c^P|T_c^I)$ can once again help properly interpret the result of the inversion. For example, profile (g) shows that for $T_c^I=1.5\times 10^6$~K, the distribution of $T_c^P$ is distributed around $T_c^P=10^7$~K, which is exactly the plasma temperature that can yield an inversion at $T_c^I=1.5\times 10^6$~K (see panel (a)). Panel (f), providing the probability distribution $T_c^P$ knowing that the inversion results is $T_c^I=7\times 10^6$~K, exhibits a broad probability distribution around $T_c^P =1.5 \times 10^{7}$, showing that the plasma temperature thus deduced is very uncertain. This is to be compared to the $0.05\ \log T_c^P$ temperature resolution of the isothermal case (see section 3.2 of Paper I). 

As the DEM becomes even larger, the impact on the robustness of the inversion becomes greater. At $\sigma^P=0.7\ \log T_e$, the probability map $P(T_c^I|T_c^P)$ of Figure~\ref{fig:multi_0.7_vs_iso} (a) and the corresponding probability $P(T_c^I)$ show clearly two privileged solutions: $T_c^I=10^6$ and $3\times 10^5$~K. The estimated isothermal temperatures are always the same for any $T_c^P$, as illustrated by panels (b) and (c). Therefore, the inversion results $T_c^I$ contains no information on the plasma central temperature $T_c^P$. This is illustrated by the lack of structure in the probability map $P(T_c^P|T_c^I)$ of panel (e). Profile (g) shows that for $T_c^I=1.5 \times 10^6$ K, the distribution of $T_c^P$ extends all over the temperature range.

\subsection{Interpretation}
\label{sec:interpretation_21}

We have shown that as the width of the plasma DEM increases, multiple solutions to the isothermal inversion appear. This phenomenon has been already mentioned by~\citet{patsourakos2007} using triple-filter TRACE data. After a proper treatment of the uncertainties, the authors found that their observations of coronal loops were consistent with both a high ($\approx 1.5\times 10^6$~K) and a low  ($\approx 5\times 10^5$~K) isothermal plasma temperature. They correctly concluded that without a priori knowledge of the physical conditions in these loops, they could not rule out the cool plasma solutions. Even though we used six bands, multiple solutions appear anyway with increasing plasma width. In addition, as we have seen in Paper I, multiple solutions can exist even with an isothermal plasma if only a limited number of bands is available. This is another illustration of the similar effects of errors and multithermality.

The isothermal temperature solutions become progressively decorrelated from the plasma central temperature as the width of the DEM increases. For very large DEMs (Figure~\ref{fig:multi_0.7_vs_iso}), the inversion process yields exclusively either $3\times 10^5$~K or $10^6$~K whatever the plasma $T_c^P$. These two temperatures correspond to the preferential locations of the minima shown in the criteria of Figure \ref{fig:criterion_gi}. This is a generalization of the phenomenon analyzed by \citet{weber2005} in the simpler case of the TRACE 19.5 over 17.3 nm filter ratio. The authors showed that in the limit of an infinitely broad DEM, the band ratio tends to a unique value equal to the ratio of the integrals of the temperature response functions. Furthermore, they showed that as the width of the DEM increases, the temperature obtained from the band ratio becomes decorrelated from the DEM central temperature. We have found a similar behavior in the more complex situation of six bands. This is not however an intrinsic limitations of AIA. We can predict that the same will occur with any number of bands or spectral lines. Indeed, for an infinitely broad DEM, since the observed intensities are equal to the product of the total emission measure by the integral of the response functions (Equation~\ref{eq:intensity_dem}), they are independent from the plasma temperature. Therefore, the inversion will yield identical results for any plasma temperature $T_c^P$, whatever the number of bands or spectral lines.

\subsubsection{Defining isothermality}
\label{chi2_near_iso}

As already noted in section~\ref{sec:ambiguity} and in Figure~\ref{criterion}, to larger DEM widths correspond larger squared residuals. From Paper I the distribution of residuals to be expected for an isothermal plasma is known. Examining then the residuals for the solutions given in the probability maps of section~\ref{sec:gauss-iso_maps}, all the solutions may not all be statistically consistent with the isothermal hypothesis. We will thus analyze the distribution of residuals to define rigorously a test of the adequacy of the isothermal model used to interpret the data.   

Because both the random and systematic errors $n_b$ and $s_b$ have been modeled by a Gaussian random variable, if the DEM model used to interpret the data can represent the plasma DEM, the residuals are equal to the sum of the square of six normal random variables (see Equation \ref{eq:criterion}). Since we adjust the two parameters $EM$ and $T_c$, the residuals should thus behave as a degree four $\chi^2$ distribution. 

\begin{figure*}
\begin{center}
\includegraphics[scale=1]{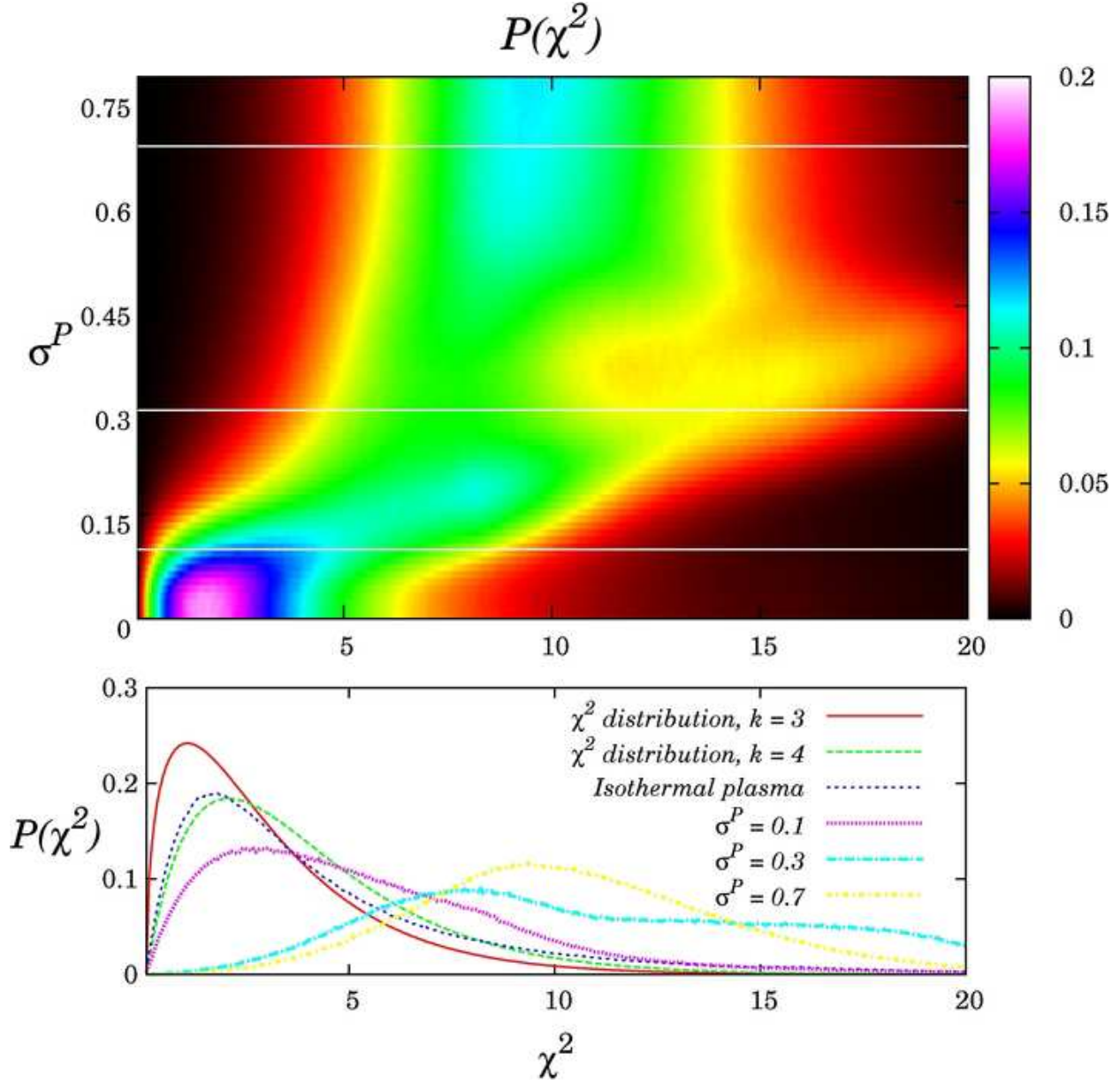}
\end{center}
\caption{In the top panel, the distributions of the sum of the squared residuals corresponding to isothermal inversions of Gaussian DEM plasmas, as a function of the DEM width $\sigma^P$. Higher probabilities correspond to lighter shades of grey. The bottom panel shows cuts at $\sigma^P=0$ (isothermal), 0.1, 0.3 and $0.7\ \log T_e$ along with theoretical $\chi^2$ distributions of degree 3 and 4.}\label{fig:chi2}
\end{figure*}

Figure~\ref{fig:chi2} shows the distribution of the squared residuals for all the 80 DEM widths considered in the simulations. The shades of grey in the top panel correspond to the probability to obtain a given $\chi^2$ value (abscissa) as a function of the width of the plasma DEM (ordinate). In the bottom panel, four profiles give the distribution of squared residuals obtained for an isothermal plasma (thin dotted line) and for the three DEM widths discussed in section~\ref{sec:gauss-iso_maps}: $\sigma^P=0.1$ (thick dotted line), 0.3 (thick dashed curve) and $0.7\ \log T_e$ (thick dash-dotted curve), corresponding to the white horizontal lines on the top panel. The theoretical $\chi^2$ distributions of degree three (thin solid curve) and four (dashed curve) are also plotted.

If the plasma is isothermal, the distribution of the residuals is slightly shifted toward a degree 3 $\chi^2$ instead of the expected degree four. In paper I, we interpreted this as a correlation between the six AIA coronal bands. The distributions of residuals progressively depart from the isothermal case as the DEM width of the simulated plasma increases. The distributions become broader and their peaks are shifted toward higher values, forming the diagonal structure in the top panel of Figure~\ref{fig:chi2}. This behavior stops around $\sigma^P = 0.4\ \log T_e$. Above, the peaks of the distributions are shifted back toward smaller values and remains constant. As the DEM becomes wider, the simulated observations become independent from the plasma parameters, and all inversions tend to give the same solution and the same residuals. 

These distributions of residuals provide a reference against which to test the pertinence of the isothermal model. The isothermal hypothesis can for example be invalidated for the solutions biased towards $T_c^I\approx 1$~MK and corresponding to very multithermal plasmas (e.g. $\sigma^P=0.7$, see section~\ref{sec:gauss-iso_maps} and Figure~\ref{fig:multi_0.7_vs_iso}). Indeed, for a given inversion and its corresponding residual, the top panel of Figure~\ref{fig:chi2} gives the most probable width of the plasma, assuming it has a Gaussian DEM. Let us assume that an isothermal inversion returns a residual equal to 5. Analyzing the histogram corresponding to the bottom row of the top panel of Figure~\ref{fig:chi2} ($\sigma^P=0$), we can show that an isothermal plasma has 68.2\% chance to yield $\chi^2\leq 5$. This residual can therefore be considered consistent with an isothermal plasma. But reading the plot vertically, we see that the probability $P(\chi^2=5)$ is greater for multithermal plasmas and peaks around $\sigma^P=0.12$, which is thus in this case the most probable Gaussian width. For larger residuals the situation is more complex because several plasma widths can have equally high probabilities. Past $\chi^2=3.5$ the plasma has a higher probability to be Gaussian than isothermal and a Gaussian inversion is required to properly determine its most probable width and central temperature. 

Figure~\ref{fig:chi2} is a generalization of the results obtained by~\citet{landi2010}. It is equivalent to their Figure 2, identifying our $\chi^2$ to their $F_\mathrm{min}$ and their solid and dashed lines to respectively the peak and the contours at half maximum of our $\chi^2$ distribution as a function of $\sigma^P$. Their Figure 2 was computed using 13 individual spectral lines for a 1~MK plasma and extends only up to $\sigma^P\approx 0.2$, while our Figure~\ref{fig:chi2} was computed for the 6 AIA bands over a wide range of central temperatures and for widths up to $\sigma^P=0.8$. Despite these differences, the two figures exhibit the same global behavior. Indeed, for any number of spectral lines or bands, the residuals of the isothermal inversion tend to increase as the width of the plasma DEM increases.

\section{Multithermal solutions}
\label{multi_response}

We now focus on multithermal solutions. The ability of AIA to reconstruct the DEM given the uncertainties is evaluated, and the probability maps associated to all parameters are computed, allowing to take into account all the Gaussian DEMs consistent with the simulated observations. Both cases of consistent and inconsistent DEMs models between the simulations and the inversion assumptions are examined. After the previous section on isothermal solutions, this generalizes the study of the impact of a wrong assumption on the DEM shape.

As in the previous section, the simulated observations remain Gaussian, but we now consider Gaussian solutions, i.e. theoretical intensities $I_{b}^{th}$ tabulated for the Gaussian DEM model. The model can thus in principle perfectly represent the plasma conditions.        

\subsection{Three-dimensional criterion}\label{sec:3D_criterion}

\begin{figure*}
\begin{center}
\includegraphics[scale=1]{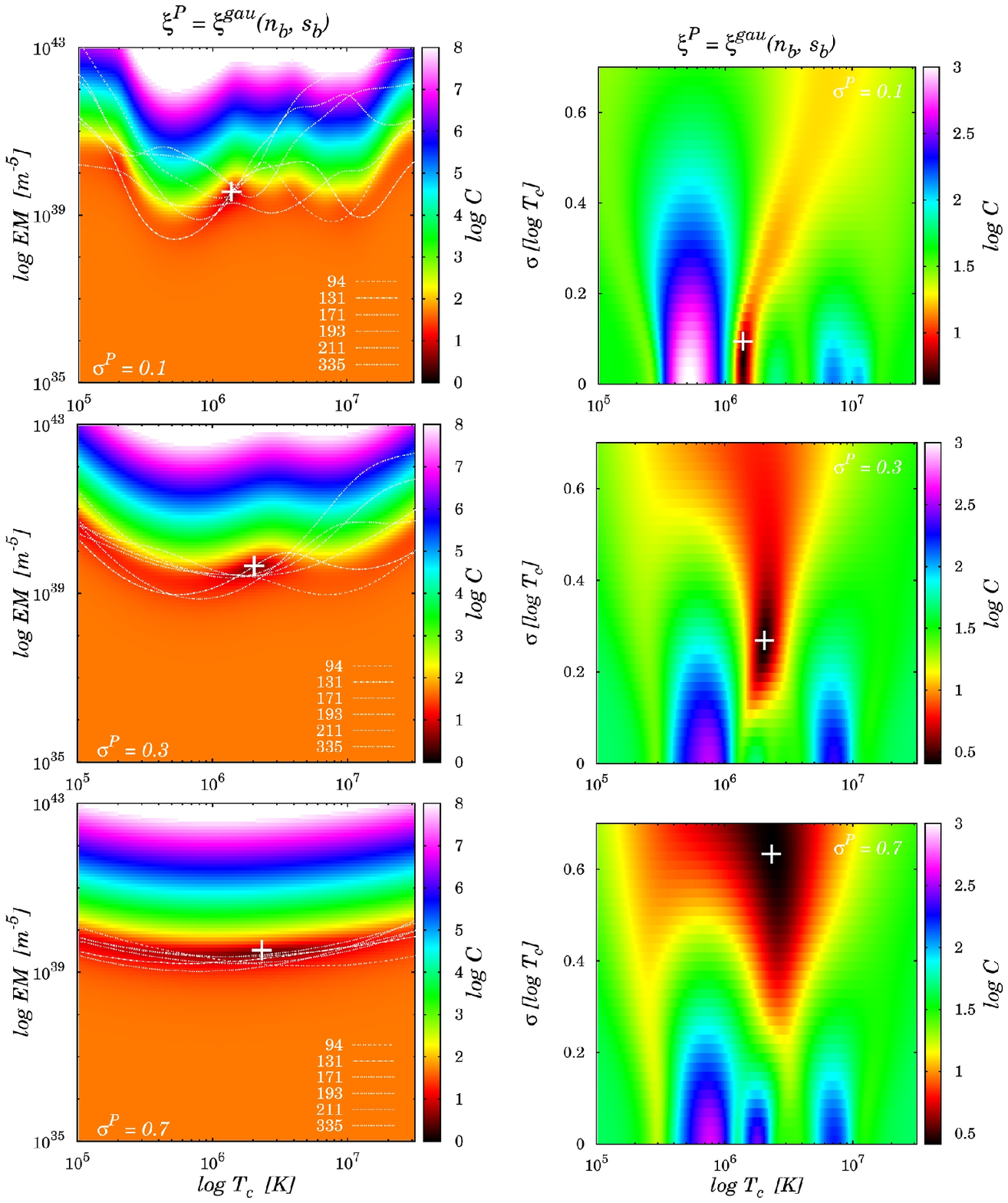}
\end{center}
\caption{Criterion in case of Gaussian multithermal DEM inversions. The three-dimensional criterion is a function of $EM^P$, $T_c^P$, and $\sigma^P$. The simulated plasma has a Gaussian DEM centered on 1.5~MK and a different width for each row. From top to bottom: $\sigma^P=0.1$, 0.3 and $0.7\ \log T_e$. The superimposed curves on the left panels represent the equivalent of the EM loci emission measure curves in a multithermal regime (see the detailed description in Section~\ref{sec:3D_criterion}).}\label{fig:crit_gvgv}
\end{figure*}

Investigating now multithermal Gaussian solutions, the least-square criterion given by Equation (\ref{eq:criterion}) has thus three dimensions $C(\xi^{gau}) = C(EM, T_c, \sigma)$. This three dimensional parameter space is systematically scanned to locate the theoretical intensities describing the best the simulated observations. Figure~\ref{fig:crit_gvgv} shows this criterion for three cases illustrating different degrees of multithermality. In each row, from top to bottom, the simulated plasma has a DEM width $\sigma^P$ of 0.1, 0.3 and $0.7\ \log T_e$ respectively, centered on the temperature $T_c=1.5 \times 10^6$~K. On the left panels, the background image represents $C(EM, T_c, \sigma^I)$, the cut across the criterion in the plane perpendicular to the DEM width axis at the width $\sigma^I$ corresponding to the absolute minimum (white plus sign). The curves are the equivalent of the loci emission measure curves in a multithermal regime: for each band $b$ and for a given DEM width $\sigma^P$, they represent the loci of the pairs $(EM, T_c)$ for which the theoretical intensities $I_b^{th}$ are equal to the observations $I_b^{obs}$. As $\sigma$ varies, they thus describe a loci \textit{surface} in the three-dimensional criterion. The difference with the isothermal loci curves is that the theoretical intensities $I_b^{th}$ have been computed for the multithermal case (i.e. considering a Gaussian DEM), and thus the parameter $\sigma$ must be now considered. The right panels of Figure \ref{fig:crit_gvgv} display $C(EM^I, T_c, \sigma)$, the cuts across the criterion in the plane perpendicular to the emission measure axis at the $EM^I$ corresponding to the absolute minimum of the criterion (also represented by a white plus sign). 

The impact of multithermality on the criterion topology is clearly visible on the loci curves, inducing a smoothing of the curves as the multithermality degree increases. Indeed, the intensities $I_b^0$ as a function of the central temperature can be expressed as the convolution product between the instrument response functions and the DEM (see Section 2.3.2 of Paper I). Therefore, the reference intensities $I_b^0(EM, T_c, \sigma)$ computed for multithermal plasmas are equal to the isothermal ones smoothed along the $T_c$ axis. As a result, the criterion exhibits a smoother topology and the minimum areas become broader and smoother as the multithermality degree increases, introducing more indetermination in the location of the absolute minimum. For each row, the realization of the uncertainties $n_b$ and $s_b$ yields a solution that is close to the simulation input. The cuts are therefore made at similar locations in the criterion in order to best illustrate the modification of its topology. We will now analyse to what extent the distortion of the criterion affects the robustness of the inversion.            
   
\subsection{Probability maps}
\label{sec:proba_gv}

In the case of DEM models defined by three parameters, and since the plasma $EM^P$ is fixed, the probability matrices $P(EM^I, T_c^I, \sigma^I|EM^P, T_c^P, \sigma^P)$ resulting from the Monte-Carlo simulations have five dimensions. In order to illustrate the main properties of these large matrices, we rely on combinations of fixed parameter values and summation over axes.

\begin{figure*}
\begin{center}
\includegraphics[scale=1]{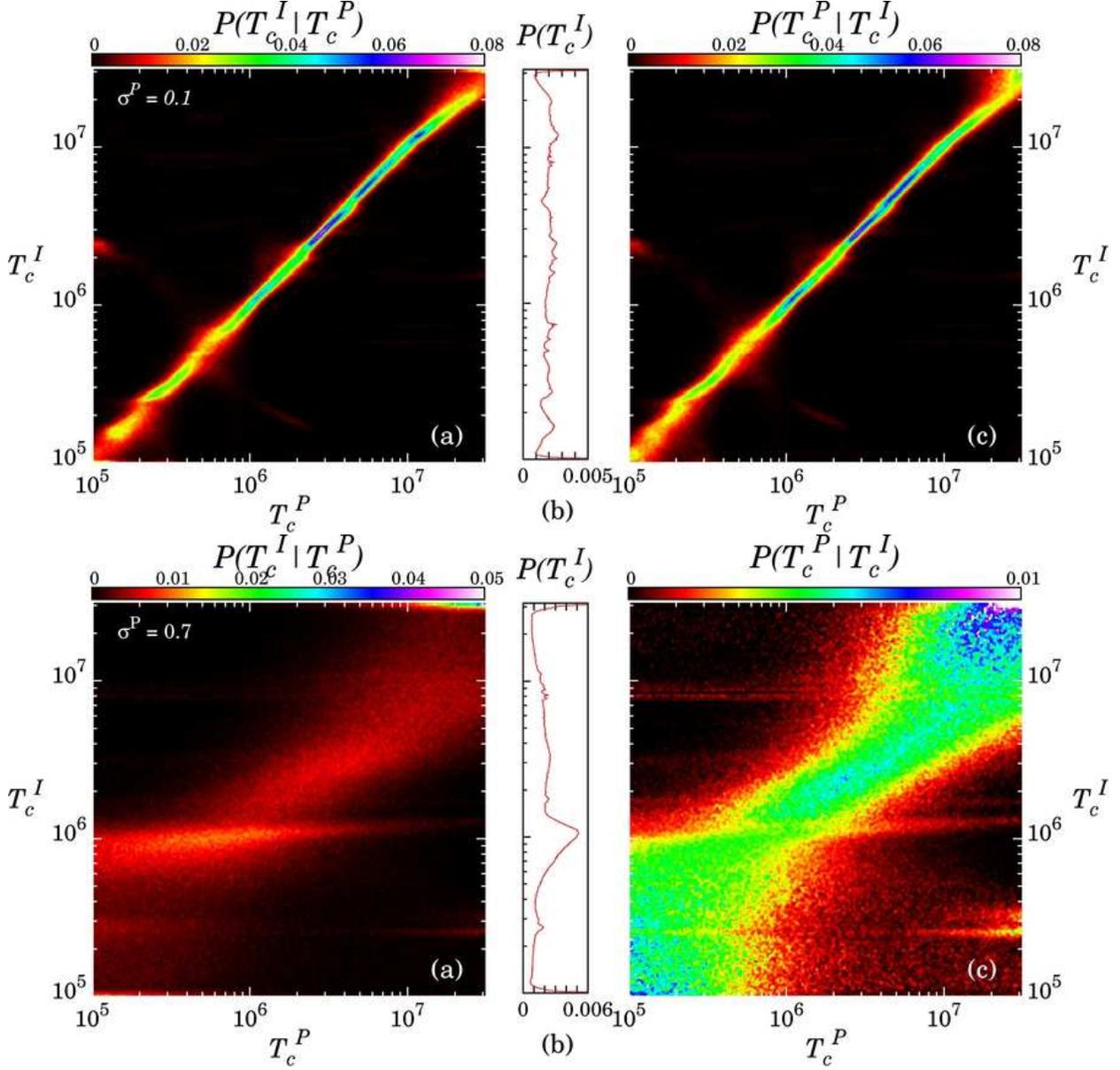}
\end{center}
\caption{Same as figure \ref{fig:multi_slight_vs_iso} but investigating now the multithermal solutions. The simulated observations are computed for DEM widths of $\sigma^P = 0.1$ (top row) and $0.7\ \log T_e$ (bottom row). The probabilities are represented here whatever the emission measure $EM^I$ and the Gaussian width $\sigma^I$ returned by the minimization scheme. In the case of low degree of multithermality plasmas, the solutions remain distributed around the diagonal. However, for broad DEM distributions, the solutions appear to be biased toward $\sim$1~MK, especially for plasmas exhibiting central temperature lower than 2~MK.}\label{fig:pro_gv}
\end{figure*}

The associated probability maps are displayed in Figure~\ref{fig:pro_gv} for simulated plasmas characterized by $\sigma^P=0.1$ (top panel) and $\sigma^P=0.7\ \log T_e$ (bottom panel)\footnote{The probability maps for 80 widths from $0$ to $0.8\ \log T_e$ are available in color on line at ftp.ias.u-psud.fr/cguennou/DEM\_AIA\_inversion/.}. The probability maps are represented whatever the emission measure and DEM width obtained by inversion, i.e. the probabilities $P(EM^I, T_c^I, \sigma^I|EM^P, T_c^P, \sigma^P)$ are integrated over $EM^I$ and $\sigma^I$. In case of a narrow DEM distribution ($\sigma^P=0.1$, top panels) the solutions are mainly distributed along the diagonal, with some secondary solutions at low probabilities. The accuracy of the determination of the central temperature $T_c^P$ is improved compared to the isothermal inversion of the same plasma as shown in Figure~\ref{fig:multi_slight_vs_iso}: the diagonal is more regular and $P(T_c^I)$ is more uniform. However, increasing the width of the DEM of the simulated plasma reduces the robustness of the inversion process. In the bottom panels ($\sigma^P=0.7$), we observe a distortion and an important spread of the diagonal. In particular, the horizontal structure in panel (a) and the consequent peak of $P(T_c^I)$ tell that the estimated temperature is biased toward $T_c^I \sim 1$~MK, especially for plasma temperatures $T_c^P < 2$~MK. This is to be compared with the privileged isothermal solutions in Figure~\ref{fig:multi_0.7_vs_iso}, and the same reasoning applies. For very broad DEMs, because of the smoothing of the $I_b^0(EM, T_c, \sigma)$ along the $T_c$ axis, the observed intensities are only weakly dependent on the central temperature, and thus all inversions tend to yield identical results. As a consequence, the probability map $P(T_c^P|T_c^I)$ shows that over the whole temperature range considered there is a large uncertainty in the determination of $T_c^P$. The smoothing of the $I_b^0(EM, T_c, \sigma)$ is due to the width of the plasma DEM and not to the properties of the response functions. It is therefore to be expected that the uncertainty in the determination of $T_c^P$ persists even if using individual spectral lines.

\begin{figure*}
\begin{center}
\includegraphics[scale=1]{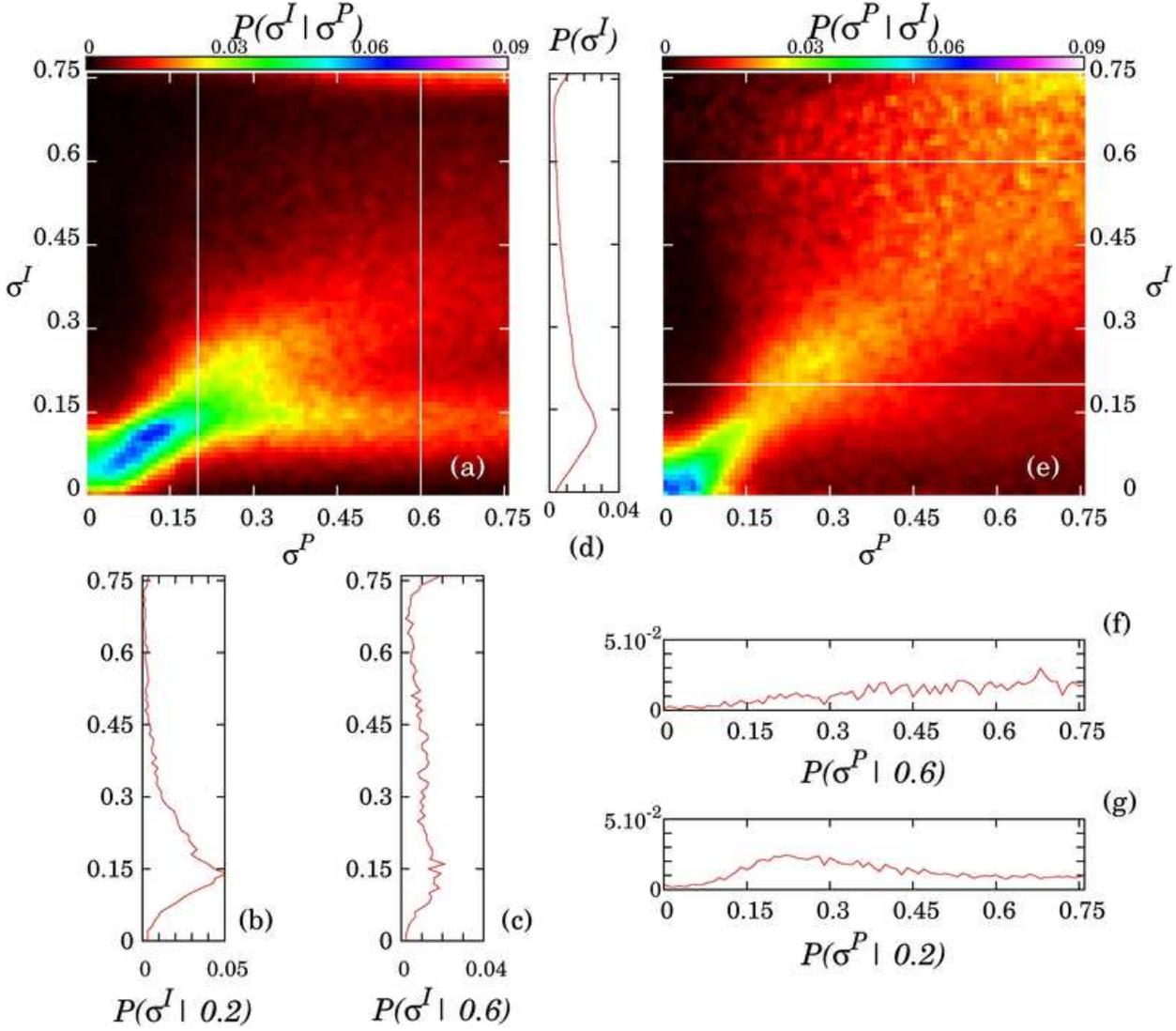}
\end{center}
\caption{Maps of probabilities for the DEM width for a simulated plasma having a Gaussian DEM centered on $T_c^P=1$~MK. Probabilities are represented whatever the emission measure $EM^I$ and the central temperature $T_c^I$ obtained. The solutions appear to be biased toward $\sigma^I \sim 0.12 \log T_e$ whatever the plasma widths, even if the situation improves for smaller widths.}\label{fig:sigma_gvgv}
\end{figure*}

In the same way, the conditional probabilities $P(\sigma^I|\sigma^P)$ and $P(\sigma^P|\sigma^I)$ of the DEM width are displayed on Figure~\ref{fig:sigma_gvgv}. The probabilities are represented whatever the estimated temperature $T_c^I$ and emission measure $EM^I$, for a simulated plasma having a DEM centered on $T_c^P=1$~MK. The conditional probability $P(\sigma^I|\sigma^P)$ on panel (a) exhibits a diagonal that becomes very wide for large $\sigma^P$ and from which another broad branch bifurcates at $\sigma^I=0.15\ \log T_e$. This branch is the analog of those observed on the temperature axis (top panels of Figure~\ref{fig:pro_gv}). For plasmas having a DEM width greater than $\sigma^P=0.3\ \log T_e$, the estimated width $ \sigma^I$ is decorrelated from the input $\sigma^P$. Profiles (b) and (c) provide the conditional probabilities of $\sigma^I$ for plasma DEM widths $\sigma^P = 0.2$ and $0.6\ \log T_e$ respectively. As shown by panel (d), the unconditional probability $P(\sigma^I)$ to obtain $\sigma^I$ whatever $T_c^I$ for this $10^6$~K plasma is biased towards $\sigma^I=0.12\ \log T_e$. Exploring the full probability matrix $P(EM^I, T_c^I, \sigma^I|EM^P, T_c^P, \sigma^P)$, we discovered that if the plasma DEM is broad, the small width solutions ($\approx 0.12\ \log T_e$) are the ones that were also biased towards a central temperature of $10^6$~K in Figure~\ref{fig:multi_0.7_vs_iso}. This is caused by the shape of the temperature response functions $R_b(T_e)$. A minimum is formed in the criterion $C(T_c, \sigma)$ around ($T_c=1$~MK, $\sigma=0.12\ \log T_e$) that tends to be deeper than the other local minima. Since for a very broad DEM plasma the simulated observed intensities $I_b^{obs}$ are almost always similar, most of the random realizations conduct to this deeper minimum and to the formation of the narrow solutions branch in $P(\sigma^I|\sigma^P)$ panel (a). Therefore, very multithermal plasmas tend to be systematically measured as near-isothermal and centered on $10^6$~K.

Reading horizontally in panel (e) the inverse conditional probability $P(\sigma^P|\sigma^I)$, a large range of DEM widths $\sigma^P$ is consistent with the estimation $\sigma^I$. Both plots (f) and (g), representing two cuts at $\sigma^I=0.2$ and $0.6\ \log T_c$, can be used to deduce the most probable plasma DEM with $\sigma^P$. However, at $\sigma^I=0.6\ \log T_e$, almost all plasma widths have significant probabilities, restricting considerably the possibility of inferring relevant DEM properties. The situation improves for smaller widths, as shown in panel (g), even though the probability distribution is still broad. For narrow DEM distributions ($\sigma^P < 0.2$), the width of the distribution decreases to about $0.15\ \log T_e$, which thus represents the width resolution limit.

The analysis of the probability maps demonstrates that the robustness of the inversion is substantially affected by the degree of multithermality of the observed plasma. Furthermore, as we already noted, the simulations presented have been made in a favorable configuration where significant signal is present in all six bands. For narrow plasma DEMs ($\sigma^P<0.15\ \log T_e$), the six AIA coronal bands enable an unambiguous reconstruction of the DEM parameters within the uncertainties. The precision of the temperature and DEM width reconstruction is then  given by the widths of the diagonals in Figures \ref{fig:pro_gv} and \ref{fig:sigma_gvgv}. For the temperature, that width is consistent with the resolution of $[0.1,0.2]\ \log T_e$ given  by \cite{judge2010}. However, both the accuracy and the precision of the inversion decrease as the multithermality degree of the simulated plasma increases, a wider range of solutions becoming consistent with the observations. In the case of a very a large DEM plasma, the solutions are skewed toward narrow 1~MK Gaussians. The isothermal solutions biased towards 1~MK (section~\ref{sec:interpretation_21}) could be invalidated on the basis of their correspondingly high residuals (section~\ref{chi2_near_iso}). But the biased Gaussian solutions are by definition fully consistent with a Gaussian plasma. This result generalizes that of~\cite{weber2005}, and since our method ensures that the absolute minimum of the criterion is found, it is a fundamental limitation and not an artefact of the minimization scheme. Instead of being evidence for underlying physical processes, the recurrence of common plasma properties derived from DEM analyses may be due to biases in the inversion processes. However, using the type of statistical analysis presented here, it is possible to identify theses biases and correct for them.

\subsection{Residuals and model testing}

\begin{figure*}
\begin{center}
\epsscale{1}
\includegraphics[scale=1]{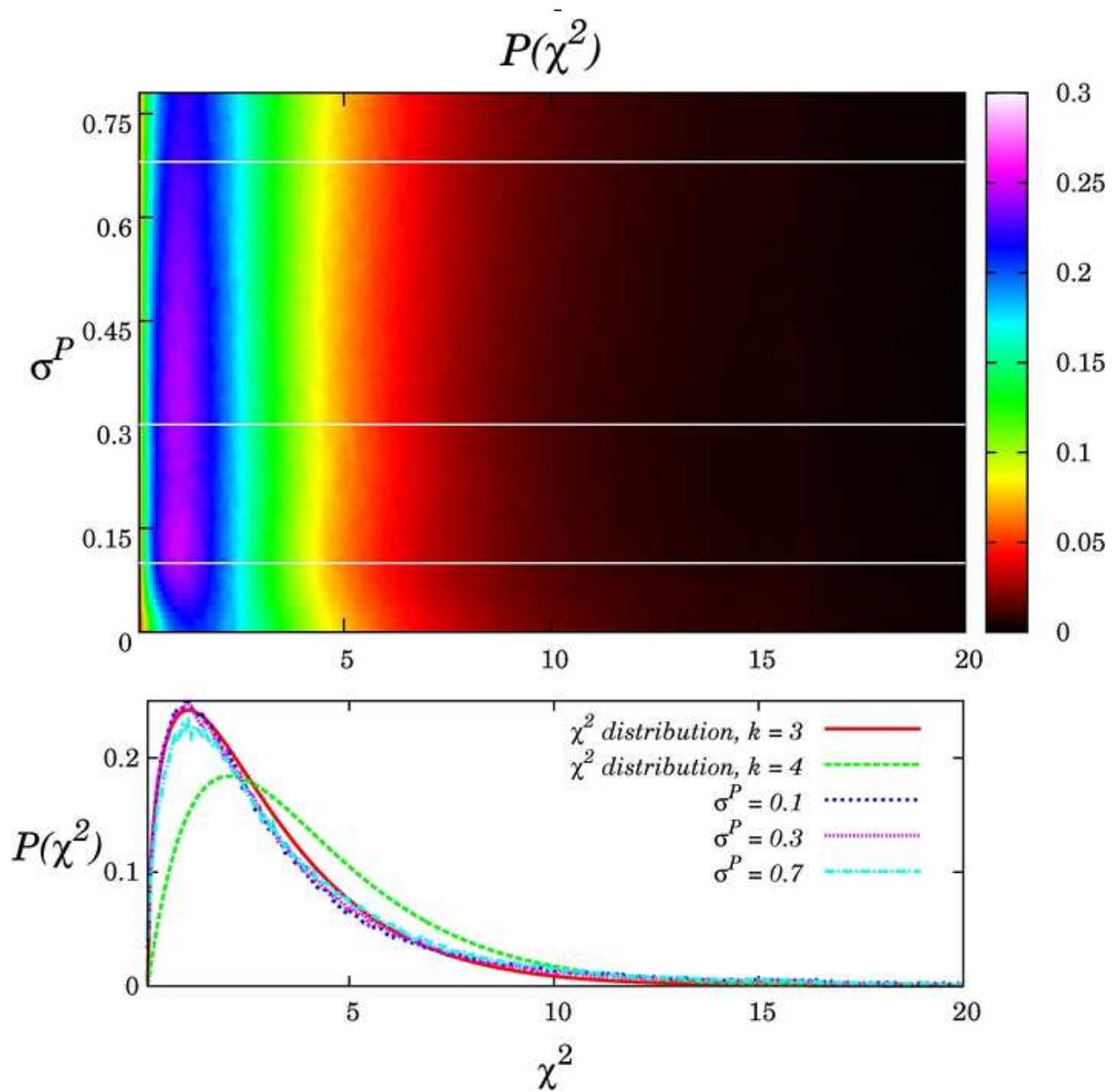}
\end{center}
\caption{Same as Figure~\ref{fig:chi2} for Gaussian solutions.}\label{fig:chi2_gv}
\end{figure*}

The distribution of the sum of squared residuals is displayed in the top panel of Figure~\ref{fig:chi2_gv} as a function of the plasma DEM width, $\sigma^P$, of the simulated plasma. The three white horizontal lines represent the locations of the cuts at $\sigma^P = 0.1$ (dotted line), 0.3 (dashed line) and 0.7 $\log T_e$ (dotted dashed line) shown in the bottom panel. Unlike for the isothermal solutions (Figure~\ref{fig:criterion_gi}), the distributions are similar for all plasma widths and resemble a degree 3 $\chi^2$ distribution (thin solid line). The most probable value is about 1 and 95$\%$ of the residuals are between 0 and 10. Any DEM inversion yielding a $\chi^2$ smaller than 10 can thus be considered consistent with the working hypothesis of a Gaussian DEM plasma. It does not mean however that a Gaussian is the only possible model, but that it is a model consistent with the observations. Since we perform a least square fit of the six values $I_b^{obs} - I_b^{th}$ by three parameters ($EM$, $T_c$, $\sigma$), this behavior is expected. The small correlation between the residuals observed in the isothermal case disappears with increasing DEM width. This can be explained by the smoothing of the criterion (see Figure~\ref{fig:crit_gvgv}) that reduces the directionality of its minima.

For model testing, a reduced  chi-squared $\chi^2_{red} = \chi^2/n$, where $n$ is the number of degree of freedom, is sometimes preferred over a regular $\chi^2$ test, for it has the advantage of being normalized to the model complexity. Usually, $n$ is the number of observations minus the number of fitted parameters. This assumes that the measurements are independent which is what we wanted to test. In practice, a small correlation between the residuals is found in the isothermal case, but the effect is small and a reduced $\chi^2$ can be used. It should be noted that the residuals have a non negligible probability to be greater than the peak of the corresponding $\chi^2$ distribution. For plasmas having Gaussian DEMs for example, the residuals have about a 43\% chance to be greater than 3 (Figure~\ref{fig:chi2_gv}). This implies that seemingly large residuals are not necessarily a proof of the inadequacy of the chosen DEM model. The working hypothesis can only be invalidated if it can be shown that another model has a greater probability to explain the obtained residuals. This was for example the case in Section~\ref{chi2_near_iso} where we have shown that if an isothermal inversion gives a squared residual greater than 3, the plasma DEM is more probably Gaussian than isothermal. We now explore a similar situation for Gaussian inversions.

Indeed, in reality the DEM shape is not known and it is interesting to test if a Gaussian is a pertinent model or if AIA has the capability to discriminate between different models. For this purpose, still considering Gaussian DEM solutions, the simulations of the observations are now performed using a top hat DEM distribution defined as    
\begin{equation}
\begin{split}
\xi_{hat}(T_e) = & EM\ \Pi_{T_c, \sigma}(T_e),\\
\mbox{with } \Pi_{T_c, \sigma}(T_e) = & \begin{cases}
                             \frac{1}{\sigma} & \text{if } |\log(T_e)-\log(T_c)| < \frac{\sigma}{2}\\
                             0       & \text{else}
                             \end{cases}
\end{split}
\end{equation}\label{eq:dem_hat}
Like a Gaussian, this parametrization can represent narrow and wide thermal structures.  

\begin{figure*}
\begin{center}
\includegraphics[scale=1]{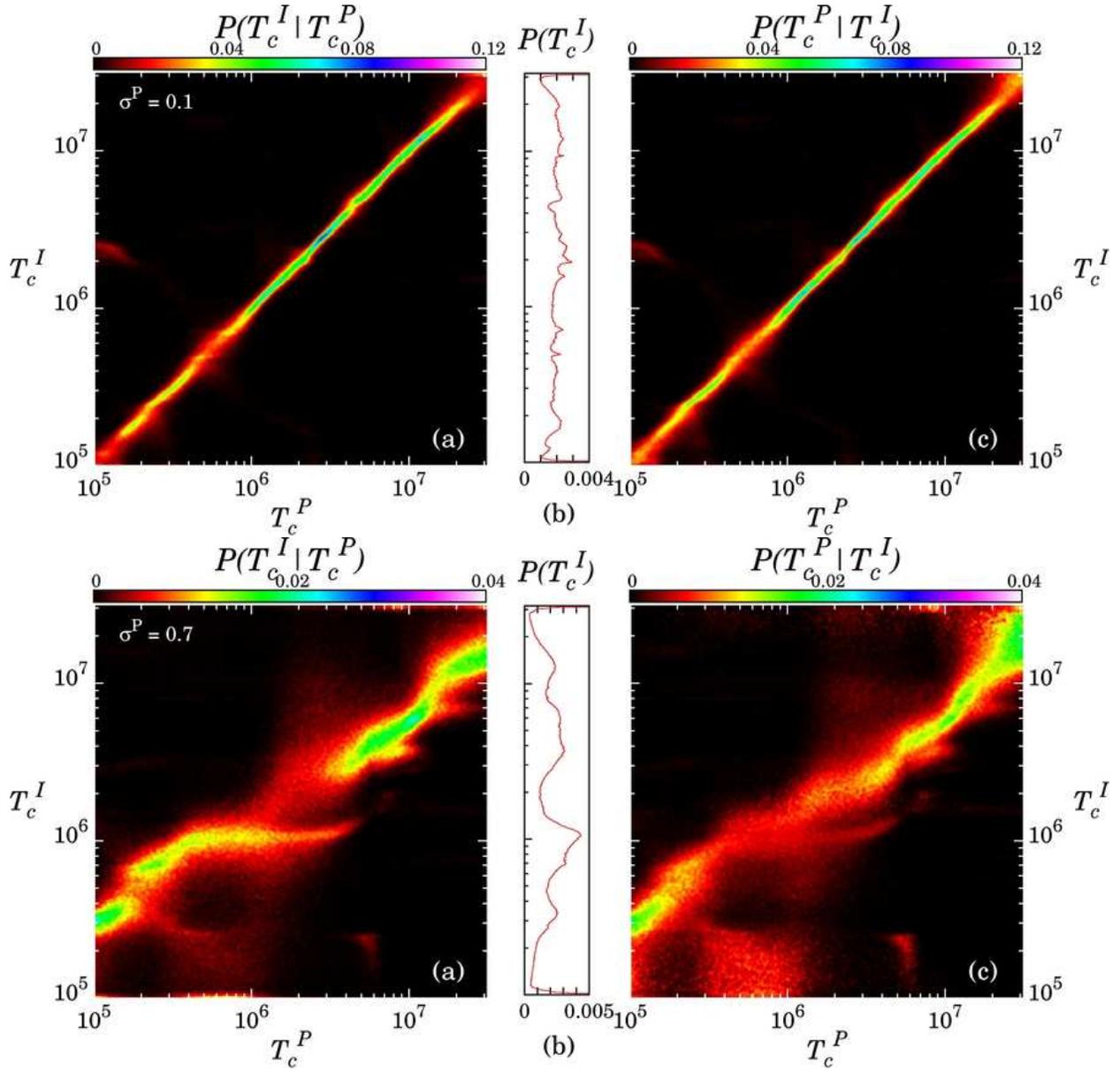}
\end{center}
\caption{Same as Figure \ref{fig:pro_gv} but for a plasma having a top hat DEM. The simulated observations are computed for DEM widths of $\sigma^P = 0.1$ (top row) and $0.7\ \log T_e$ (bottom row).}\label{fig:pro_egv}
\end{figure*}

Figure~\ref{fig:pro_egv} gives the associated temperature probability maps. The top panels correspond to simulated plasmas with a top hat distribution of width $\sigma^P=0.1$. Most of the solutions are concentrated around the diagonal, even though the robustness is somewhat affected for temperatures in the range $5 \times 10^5 <T_c^P < 10^6$~K, where low probability secondary solutions exist. These two plots are very similar to the top panels of Figure~\ref{fig:pro_gv} that give the probability of the Gaussian solutions to a Gaussian plasma. Therefore, even though the DEM assumed for the inversion is different from that of the plasma, the central temperature of the top-hat is determined unambiguously.

The picture is different for wide plasma DEM distributions. In the bottom panels of Figure \ref{fig:pro_egv}, with $\sigma^P = 0.7$, the diagonal has become very wide, structured and does not cross the origin any more. Some plateaus appears, meaning that for some ranges of plasma temperature $T_c^P$, the temperature $T_c^I$ provided by the Gaussian solutions will be invariably the same. For $5 \times 10^5 < T_c^P < 3 \times 10^6$~K for example,  a constant solution $T_c^I = 1$~MK appears, as shown also by $P(T_c^I)$ (bottom panel (b)). As a result, the inverse conditional probability map $P(T_c^P|T_c^I)$ (bottom panel (c)) indicates that for an inversion output of 1~MK there is a large indetermination on the central temperature. The behavior of the solutions is thus globally equivalent to that described in Section~\ref{sec:proba_gv} for the inversion of Gaussian DEM plasmas.

The observed distribution of the sum of the squared residual is very similar to those obtained with consistent Gaussian DEM models and is close to 3 $\chi^2$ (see Figure ~\ref{fig:crit_gvgv}). If the working hypothesis were a top hat DEM, thus consistent with the plasma, the residuals would also be close to a degree 3 $\chi^2$, as for any DEM model described by three parameters (see discussion above). The solar plasma has no reason to really have a top hat DEM. However, this numerical experiments shows that, using AIA data only and the $\chi^2$ test, the discrimination between two very different multithermal DEMs is practically impossible. 

\section{Summary and discussion}
\label{discussion}

In this work we described a complete characterization of the statistical properties of the DEM inversion, rigorously treating both systematic and random errors. The developed methodology has been illustrated in the specific case of the AIA telescope, but the technique is generic and can be applied to any other instrument, spectrometers as well as imaging telescope. By restricting ourselves to parametric DEMs, we could analyze in details what occurs during the inversion process, and could therefore point out the fundamental difficulties involved in the DEM reconstruction. Even though only a few simple DEM distributions have been studied, important and generic conclusions regarding the robustness of the inversion problem have been reached. The method can be applied to other forms of DEMs as long as they can be defined by a small number of parameters.

Our technique provides new tools to facilitate the interpretation of the DEM inversion. By computing the $P(\xi^I|\xi^P)$ probability matrices the robustness can be evaluated, secondary solutions can be detected and their probabilities quantified. Since we do not know whether the systematic errors are over or under-estimated, their randomization in the computation of the inverse conditional probability $P(\xi^P|\xi^I)$ ensures that all possible DEMs consistent with the solution $\xi^I$ given the uncertainties are taken into account. The specific location of secondary solutions in the space of the possible DEMs $\xi$ can be interpreted by analyzing the topography of the criterion $C(\xi)$, which is intrinsically linked to the shapes of the temperature response functions. The simulated distributions of the sum of the squared residuals provide references against which to asses the pertinence of the DEM model used to perform the inversion. Applying the technique on real data, and applying the tools presented here on the results, the interpretation of DEM inversions can be facilitated.      

By investigating the isothermal solutions, we have shown that the six AIA coronal bands greatly enhance the robustness of the DEM reconstruction compared to previous EUV imagers, even though the temperature resolution varies between 0.03 and $0.11\ \log T_c$. This resolution depends on the level of uncertainties assumed (see Section 3.2 in Paper I). Because of the ambiguity between measurement errors and deviation from isothermality, a variety of plasma conditions is consistent with the isothermal hypothesis. As the degree of multithermality of the observed plasma increases, the isothermal solutions become progressively decorrelated from the input. We have demonstrated that, in the limit of very broad multithermal plasmas, the isothermal solutions are biased toward two preferred solutions: $3\times 10^5$ or $10^6$~K depending on the systematics. But the analysis of the behavior of the residuals allows to measure the adequacy of the isothermal hypothesis. Past a critical value of the residuals, a multithermal DEM is more probable than an isothermal plasma. That critical value has to be estimated from Monte-Carlo simulations for the considered set of bands or spectral lines. Our procedure provides a rigorous quantification of the classical EM loci isothermal test.     
 
Investigation of the multithermal Gaussian solutions shows that AIA has the ability to reconstruct simple DEM forms. However, the greater the width of the plasma DEM, the lower the accuracy and precision on the determination of the DEM parameters.  We argued that this would be the case also if using individual spectral lines instead of narrow bands. In the extreme case of very broad DEMs distributions the quality of the inversion is strongly affected, because of the combined effects of a smoothed criterion and simulated observations weakly dependent on the plasma parameters. The AIA Gaussian solutions are biasS.P. acknowledges the support from the Belgian Federal Science Policy Office through the international cooperation programmes and the ESA-PRODEX programme and the support of the Institut d'Astrophysique Spatiale (IAS). F.A. acknowledges the support of the Royal Observatory of Belgium. The authors would like to thank J. Klimchuk for fruitful discussions and comments.ed towards a near isothermal plasma ($\sigma^I = 0.15$) centered on $T_c = 1$~MK. Furthermore, the analysis of the distribution of the residuals has revealed that it is difficult to discriminate between different multithermal DEMs models using AIA alone.

These results can have far reaching implications for the determination of the thermal structure of the solar corona. Indeed, the thermal structure of coronal loops for example is still under debate, different proposed heating scenarios leading to opposite DEM properties. For example, the nanoflares theory \citep[e.g.][]{parker1983, cargill1994, klimchuk2006}, suggesting that the stored energy is then released impulsively over small scales, predicts a wide thermal structure at a given time, in case of sparse events inside a single strand. Conversely, the steady heating processes should produce loops with narrow DEMs. Our analysis revealed that even if the residuals support the pertinence of the solution, the DEM analysis of AIA observations can lead to the erroneous conclusion that the plasma is nearly isothermal while it is in fact broadly multithermal. Different biases can be expected for different sets of bands or spectral lines, different instruments or atomic physics data.

The methodology developed in this series of papers can also be used to establish the optimal set of bands needed to maximize the robustness of the DEM inversion, for broad band instrument as well as spectrometers. Performing this type of preparatory simulations can thus be very useful to optimize the design of future solar instruments if they are to provide reliable temperature diagnostics.   

\acknowledgments
S.P. acknowledges the support from the Belgian Federal Science Policy Office through the international cooperation programmes and the ESA-PRODEX programme and the support of the Institut d'Astrophysique Spatiale (IAS). F.A. acknowledges the support of the Royal Observatory of Belgium. The authors would like to thank J. Klimchuk for fruitful discussions and comments.

\acknowledgments

\clearpage

\clearpage

\clearpage

\clearpage

\clearpage

\clearpage

\clearpage

\clearpage

\clearpage

\clearpage

\clearpage

\clearpage

\end{document}